# Linear Least Squares Estimation of Fiber-Longitudinal Optical Power Profile

Takeo Sasai, *Member, IEEE*, Minami Takahashi, Masanori Nakamura, *Member, IEEE,* Etsushi Yamazaki, *Member, IEEE,* and Yoshiaki Kisaka

*Abstract*—This paper presents a linear least squares method for fiber-longitudinal power profile estimation (PPE), which estimates an optical signal power distribution throughout a fiber-optic link at a coherent receiver. The method finds the global optimum in least square estimation of longitudinal power profiles, thus closely matching true optical power profiles and locating loss anomalies in a link with high spatial resolution. Experimental results show that the method achieves accurate PPE with an RMS error from OTDR of 0.18 dB. Consequently, it successfully identifies a loss anomaly as small as 0.77 dB, demonstrating the potential of a coherent receiver in locating even splice and connector losses. The method is also evaluated under a WDM condition with optimal system fiber launch power, highlighting its feasibility for use in operations. Furthermore, a fundamental limit for stable estimation and spatial resolution of least-squares-based PPE is quantitatively discussed in relation to the ill-posedness of PPE by evaluating the condition number of a nonlinear perturbation matrix.

## I. Introduction

CHARACTERIZING physical parameters of transmission links is essential to reduce redundant operational margins [1], identify soft network failures [2], and construct digital-twins of optical networks [3]. Various monitoring methods have been proposed to estimate link parameters, such as optical signal-to-noise ratio (OSNR), fiber nonlinearity, and chromatic dispersion (CD), using receiver-side (Rx) digital signal processing (DSP) [4]. These Rx-DSP-based approaches are cost-effective compared to dedicated hardware-based approaches; however, they typically estimate cumulative parameters of the entire link. Consequently, the spatial resolution of estimated parameters is often limited, making the pinpointing of network faults challenging. If parameters distributed in the fiber-longitudinal direction are obtained in Rx DSP, an intelligent transponder can be expected that not only accurately predicts the achievable rate of a link but also locates soft failures without dedicated hardware devices, thereby reducing operational costs.

Given this context, studies on digital longitudinal monitoring of fiber-optic links have emerged [5]-[18]. This approach employs a coherent Rx DSP to monitor various link parameters longitudinally distributed along fibers, such as signal power profiles [5]-[12], span-wise CD maps or fiber types [5], [9], [10], gain spectra of individual amplifiers [5], [13], [14], responses of individual optical filters [5], [11], the location of excessive polarization dependent loss [15][16][17], and multi-path interference [18]. The advantages of DLM lies in its capability of characterizing an entire multi-span link and locating anomalous link components without the need for dedicated hardware devices such as optical time domain reflectometry (OTDR) and optical spectrum analyzers. This is particularly beneficial in disaggregated network scenarios [19], where network elements of multiple vendors and domains coexist and the link parameters are not always shared. The DLM offers a solution to this issue by monitoring link parameters end-to-end using a single coherent receiver, regardless of the availability of link parameters from different vendors and domains. Using this approach, a dynamic optical path provisioning for an end-to-end connection over links with unknown access domains was demonstrated in [20].

Among the parameters monitored in DLM, fiber-longitudinal power profile estimation (PPE) [5]-[12] is of particular importance due to its utility in estimating fiber nonlinear interference and locating loss (and gain) anomalies. PPE obtains the fiber-longitudinal optical power profile by estimating distance-wise nonlinear phase rotations $\gamma'(z) = \gamma(z)P(z)$ from Rx signals, where $\gamma(z)$ and $P(z)$ are the nonlinear constant of a fiber and the signal power at position $z$ on the fiber, respectively. The estimation of $\gamma'(z)$ can be seen as an inverse problem of the nonlinear Schrödinger equation, where its coefficient $\gamma'(z)$ is estimated from boundary conditions (i.e., Tx and Rx signals) [6]. In general, a common strategy for solving such an inverse problem is the least squares (LS) method, where parameters are estimated as the optimum values that minimizes square errors. However, PPE is a *nonlinear* least squares problem, which makes finding the global optimum challenging. The gradient descent optimization of the split-step method [5], [9] is a straightforward approach, but it requires iterative optimization and careful selection of various hyperparameters such as the learning step size, number of iterations, and optimizer; otherwise, the estimated power profiles would be easily trapped into a local minimum, thereby limiting the measurement accuracy of PPE. Although correlation-based methods (CMs) [7], [8], [18] can avoid the iterative optimization, it has been shown in [6] that CMs inherently have limited accuracy and spatial resolution even under noiseless and distortionless conditions. Additionally, CMs do not estimate the true value of $\gamma'(z)$ without hardware-based calibration [21].

In this paper, we propose and experimentally demonstrate a *linear* least squares method for PPE that finds the global optimum in the least square estimation of the nonlinear phase $\gamma'(z)$. The method achieves a high measurement accuracy and spatial resolution in estimating the true value of $\gamma'(z)$. Experimental results show that the proposed method achieves agreement with OTDR results with an RMS error of 0.18 dB without any calibration. Consequently, a loss anomaly of 0.77 dB in a 50-km × 3-span link is successfully located, demonstrating that a coherent receiver can locate even poor splices or connectors in multi-span links as OTDR does. We first show that the nonlinear least squares problem of estimating $\gamma'(z)$ can be reduced to a linear least squares problem by applying the first order regular perturbation (RP1) model. We then numerically demonstrate that power profiles estimated by the linear least squares aligns well with the theoretical power profiles, locating even a 0.2-dB loss anomaly. A fundamental limit for stable estimation and spatial resolution of the LS-based PPE is also quantitatively discussed by evaluating the ill-posedness of PPE. We finally present our experimental results under both ideal and practical configurations, including with and without WDM channels, as well as high and optimum fiber launch powers.

This paper extends the work presented in [12], [22] with the following additional discussions:
- The connection between CMs and linear least squares.
- The fundamental limit for stable estimation of LSs in relation to the ill-posedness of PPE.
- The achievable spatial resolution of LSs.
- Experimental results under practical link conditions with system optimal launch power and WDM channels.

The remainder of this paper is organized as follows. Section II details the problem formulation, the algorithm, and simulation results as well as limitations of the proposed linear least squares. Section III presents experimental results under ideal conditions with high fiber launch power, and under practical conditions with low power and WDM conditions. Section IV concludes the paper.

## II. LINEAR LEAST SQUARES FOR PPE

### A. Problem Formulation

PPE can be formulated as an inverse problem of the nonlinear Schrödinger equation (NLSE), where the nonlinear coefficients are reconstructed from boundary conditions, i.e., Tx and Rx signals. The propagation of the optical signals $A \equiv A(z,t)$ in optical fibers at position $z \in [0, L]$ and time $t$ is governed by NLSE:

$$\frac{\partial A}{\partial z} = \left( j\frac{\beta_2(z)}{2}\frac{\partial^2}{\partial t^2} + \frac{\beta_3(z)}{2}\frac{\partial^3}{\partial t^3} \right) A - j\gamma'(z)|A|^2 A, \quad (1)$$

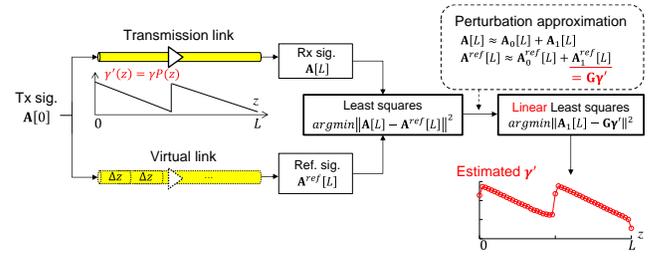

Fig. 1. Conceptual configuration of the proposed linear least squares for PPE. In this work, linear least squares estimation is performed between Rx and reference signals to estimate longitudinal power profile $\propto \gamma'(z)$. Both signals are approximated by regular perturbation model.

$$\gamma'(z) \equiv \gamma(z)P(0)\exp\left(-\int_0^z \alpha(z')\,dz'\right) = \gamma(z)P(z), \quad (2)$$

where $\alpha(z), \beta_2(z), \beta_3(z), \gamma(z)$, and $P(z)$ are the fiber loss, second/third dispersion, nonlinear constant, and optical signal power at $z$, respectively. Note that, in this formulation, $\alpha(z)$ is merged into the nonlinear coefficient $\gamma'(z)$, and the power of $A$ is normalized to 1 regardless of position $z$ [5]. In turn, all the power variation due to the fiber loss and amplification is governed only by $\gamma'(z)$. Thus, $P(z)$ can be inferred by estimating $\gamma'(z)$, assuming $\gamma(z)$ is constant. For simplicity, single polarization is assumed here and in subsequent simulations. In experiments, the algorithm is extended to dual polarization, based on the Manakov equation instead of NLSE (see Appendix A).

In this work, $\gamma'(z)$ is estimated as the optimal nonlinear coefficients that best reproduce the received signals. Fig. 1 shows the conceptual configuration of the proposed method. The transmitted signals $\mathbf{A}[0] = [A(0,0), \dots, A(0,nT), \dots, A(0,(N-1)T)]^T$ are launched into a fiber link governed by the NLSE (upper tributary) and evolve into $\mathbf{A}[L]$ at a coherent receiver, where $n \in [0, N-1]$ is the time sample number, $T$ is the sampling period, $(\cdot)^T$ is a transpose, and $L$ is the total distance. On the other tributary, $\mathbf{A}[0]$ also propagates a virtual link that emulates the fiber link in the digital domain and evolves into a reference $\mathbf{A}^{ref}[L]$. Several models for the virtual link have been proposed such as the full split-step method (SSM) [5], [9], simplified SSM [7], [8], simplified SSM with nonlinear template [18], and Volterra series expansion [11]. In this work, the virtual link is modeled by RP1 [23], [24]. By doing so, an analytical expression for the estimation of $\gamma'(z)$ is obtained. The estimation of $\gamma'(z)$ can be formulated as a classical least squares problem:

$$\begin{aligned}\widehat{\boldsymbol{\gamma}'} &= \underset{\boldsymbol{\gamma}'}{\operatorname{argmin}}\, I \\ &= \underset{\boldsymbol{\gamma}'}{\operatorname{argmin}} \left\| \mathbf{A}[L] - \mathbf{A}^{ref}[L] \right\|^2 \end{aligned} \quad (3)$$

where $\boldsymbol{\gamma}' = [\gamma'_0, \gamma'_1, \dots, \gamma'_{K-1}]^T$, and $\gamma'_k$ is the discretized version of $\gamma'(z)$ at position $z_k$ ($k \in \{0, \dots, K-1\}$, and thus $z_{K-1} = L$). $\mathbf{A}^{ref}[L]$ is the emulated Rx signal after propagating through the virtual link and is a function of $\gamma'_k$. The problem (3) is a *nonlinear* least squares problem since $\gamma'_k$ appears in exponential functions for nonlinear phase rotations. However, (3) can be reduced to a *linear* least squares problem by using RP1 as described in the following subsection. The essence is that, in RP1, the perturbation signal vector is expressed as a linear system of equations (see (7)), which guarantees the global optimal solution of (3) expressed in an analytical form (11).

### B. Derivation of Linear Least Squares

In this work, both $\mathbf{A}[L]$ and $\mathbf{A}^{ref}[L]$, $N \times 1$ vectors, are modelled by using RP1 such that

$$\mathbf{A}[L] \simeq \mathbf{A}_0[L] + \mathbf{A}_1[L], \tag{4}$$

$$\mathbf{A}^{ref}[L] \simeq \mathbf{A}_0^{ref}[L] + \mathbf{A}_1^{ref}[L], \tag{5}$$

where $\mathbf{A}_0$ and $\mathbf{A}_0^{ref}$ are the linear terms obtained by applying CD to the Tx signals, while $\mathbf{A}_1$ and $\mathbf{A}_1^{ref}$ are the first-order perturbation terms. Note that $[L]$ is removed for simplicity unless specified. Then the cost function $I$ in (3) becomes

$$\begin{aligned} I &\simeq \left\| (\mathbf{A}_0 + \mathbf{A}_1) - (\mathbf{A}_0^{ref} + \mathbf{A}_1^{ref}) \right\|^2 \\ &= \left\| \mathbf{A}_1 - \mathbf{A}_1^{ref} \right\|^2 \end{aligned} \tag{6}$$

where $\mathbf{A}_0 \simeq \mathbf{A}_0^{ref}$ is assumed since the linear term $\mathbf{A}_0$ can be well approximated in the digital domain by applying digital filters for CD [25], [26]. $\mathbf{A}_1^{ref}$ is explicitly expressed in RP1 as:

$$\mathbf{A}_1^{ref} = \mathbf{G}\boldsymbol{\gamma}', \tag{7}$$

where

$$\mathbf{G} = [\boldsymbol{g}_0, \dots, \boldsymbol{g}_k, \dots, \boldsymbol{g}_{K-1}], \tag{8}$$

$$\boldsymbol{g}_k = -j\Delta z_k \mathbf{D}_{z_k L} \widetilde{\mathbf{N}} \left[ \mathbf{D}_{0z_k} \mathbf{A}[0] \right]. \tag{9}$$

Here, $\Delta z_k = z_{k+1} - z_k$, $\mathbf{D}_{z_1 z_2}$ is a matrix for CD from $z_1$ to $z_2$, and $\widetilde{\mathbf{N}}[\cdot] = (|\cdot|^2 - 2\bar{P})(\cdot)$ is a nonlinear operator with element-wise multiplication, where $\bar{P}(= 1)$ is the power of $\mathbf{E}$. By substituting (7) into (6), the cost function becomes

$$I \simeq \|\mathbf{A}_1 - \mathbf{G}\boldsymbol{\gamma}'\|^2 \tag{10}$$

This can be solved by linear least squares. Considering that $\boldsymbol{\gamma}'$ is a real vector, the solution is as follows (see Appendix B for the derivation):

$$\widehat{\boldsymbol{\gamma}'} = (\text{Re}[\mathbf{G}^\dagger \mathbf{G}])^{-1} \text{Re}[\mathbf{G}^\dagger \mathbf{A}_1]. \tag{11}$$

This is the proposed linear least squares algorithm. In the following, the spatial step size is assumed to be uniform as $\Delta z_k = \Delta z$.

*Remark 1:* Instead of taking the real parts as in (11), Kim et al. proposed a simpler form $\widehat{\boldsymbol{\gamma}''} = (\mathbf{H}^\dagger \mathbf{H})^{-1} \mathbf{H}^\dagger \mathbf{A}$ by augmenting the matrix $\mathbf{G}$ and coefficient vector $\boldsymbol{\gamma}'$ as $\mathbf{H} = [\mathbf{G} \ \mathbf{A}_0]$ and $\boldsymbol{\gamma}'' = c[\boldsymbol{\gamma}'^T \ 1]^T$, where $c$ is a complex-valued scaling factor [27], [28]. This modification ensures that the estimated $c$ automatically compensates for a phase mismatch between a reference $\mathbf{A}^{ref}$ and Rx signals after carrier phase recovery $\mathbf{A}$, enhancing the robustness of PPE to link conditions.

### C. Relation to Correlation-based Methods

As another PPE method, CMs have been proposed [7], [8], [18]. In a modified CM originally proposed in [18] and detailed in [6], a reference $\mathbf{A}^{ref}[L]$ is obtained by applying the CD, nonlinear operator, and residual CD to Tx signals. Notice that this operation is the same as $\boldsymbol{g}_k$ in (9). The power at $\boldsymbol{z}_k$ is then estimated from the correlation between this reference and Rx signals as $\text{Re}[\boldsymbol{g}_k^\dagger \mathbf{A}]$. This process is iterated for all positions $\boldsymbol{z}_k$ to construct a power profile as $\text{Re}[\mathbf{G}^\dagger \mathbf{A}]$. Under the assumption that signals in fibers are a stationary Gaussian process, the resulting power profile was analytically shown [6] to be a convolution between a true power profile and a smoothing function originating from the spatial correlation of the nonlinearity. Due to this convolution effect, the spatial resolution and measurement accuracy of CMs are limited, as shown in the subsequent simulation. To address this issue, Hahn et al. applied a deconvolution of such a smoothing function to enhance spatial resolution [29].

Interestingly, CMs have a close relationship to the linear least squares presented in this work. According to [6], the correlation for the linear part, $\text{Re}[\mathbf{G}^\dagger \mathbf{A}_0]$, is zero under the Gaussian signal assumption. Therefore, the estimated power profile of CMs can be reduced to $\text{Re}[\mathbf{G}^\dagger \mathbf{A}] = \text{Re}[\mathbf{G}^\dagger \mathbf{A}_1]$. Notice that this expression also appears in the derived linear least squares (11). This suggests that there is a strong connection between the linear least squares and CMs, and the primary distinction is the presence of an inverse $(\text{Re}[\mathbf{G}^\dagger \mathbf{G}])^{-1}$ [30]. This inverse matrix cancels the convolution effect in CMs, allowing the linear least squares to achieve high spatial resolution and measurement accuracy.

*Remark 2:* What is the difference in applying a simple deconvolution and the inverse matrix to CMs? Applying a deconvolution is a special case of applying the inverse matrix (i.e., linear least squares). If signals satisfy the stationary Gaussian assumption, these operations coincide since $\text{Re}[\mathbf{G}^\dagger \mathbf{G}]$ becomes a Toeplitz matrix (i.e., a linear convolution) and its inverse provides a deconvolution [6]. However, if signals do not follow a stationary Gaussian process, CMs can no longer be expressed as a convolution [6]. Instead, the convolution effect of CMs becomes position-dependent, which cannot be fully canceled by a simple deconvolution. For instance, with practical modulation formats such as QPSK and 16QAM, CMs exhibit weaker power than expectation at the beginning of the link [6], [31], [32] as will be shown in Fig. 2(a), which means the convolution effect is position-dependent and modulation-format-dependent. To fully remove such a position-dependent convolution, one should apply the inverse matrix. By doing so, the excessively weak power in CMs is corrected in the least squares, which also implies that the LS does not show significant modulation format dependency [6]. In summary, $(\mathrm{Re}[\mathbf{G}^{\dagger}\mathbf{G}])^{-1}$ serves as a generalized form of deconvolution, applicable to a wide range of modulation formats.

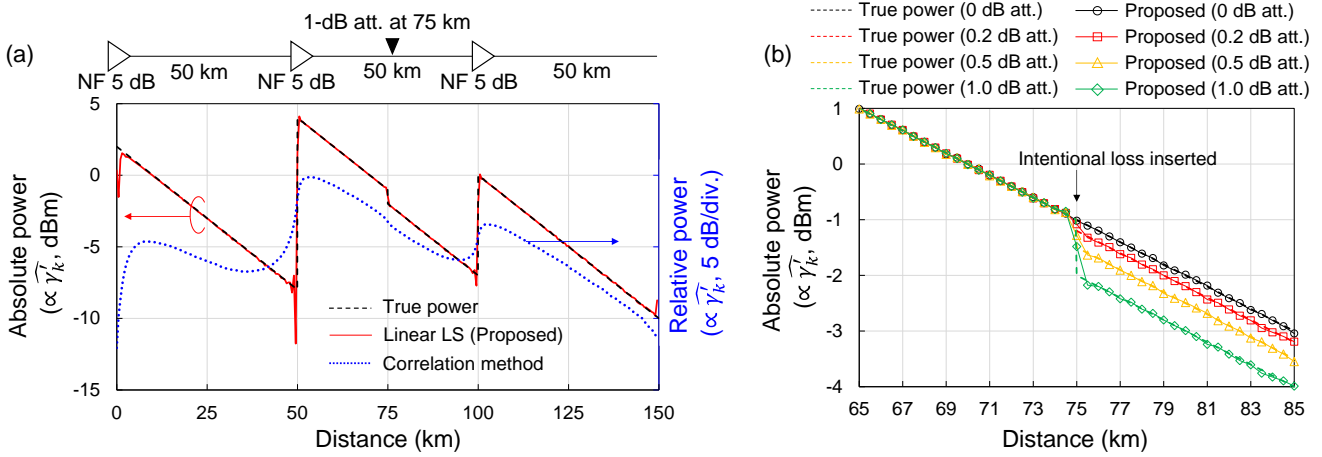

Fig. 2. (a) Simulation results of PPE for 50-km × 3-spans link using proposed linear least squares (red) and correlation method (CM, blue) with 1.0-dB intentional attenuation inserted at 75 km. (b) Magnified version around inserted loss with various attenuation levels. 16QAM 128GBd signals were used.

*D. Other Related Works*

In our experiments, we leverage nonlinear self-channel interference (SCI) including cross-polarization phase modulation (XPolM) to estimated power profiles as described in Appendix A. Another demonstration of estimating longitudinal nonlinear phases and span-wise CDs using XPolM between orthogonally polarized special tones can be found in [33]. The authors of [34] leveraged cross-channel interference (XCI) between WDM channels to locate a loss anomaly. This XCI-based approach achieves higher spatial resolution than SCI-based methods due to a larger walk-off between interfering channels.

A machine-learning-based method was also introduced to identify fiber coefficients in NLSE from observation data [35]. While the estimated coefficients were constant and not longitudinally distributed, the technique was then applied to predict a power evolution and the Raman gain spectrum in the C+L band [36]. Even though anomaly detection was not part of their demonstration, these approaches can be applied to QoT estimation for optical path provisioning.

*E. Simulation*

A 16QAM 128-GBd signal with a root-raised-cosine (RRC) roll-off factor of 0.1 was generated and launched into the link with a 50-km × 3-span. Lumped amplifications were placed at the beginning of the spans with a noise figure (NF) of 5.0 dB. To emulate fiber propagation, the split-step Fourier method was used with a spatial step size of 50 m and an oversampling ratio of 8 samples/symbol. The fiber parameters were $\alpha$ = 0.20 dB/km, $\beta_2$ = -21.6 ps$^2$/km, and $\gamma$ = 1.30 W$^{-1}$km$^{-1}$. A single polarization transmission was assumed. After downsampled to two samples/symbol, the signal experiences CD compensation, synchronization, and CD reloading. A perturbation vector $\mathbf{A}_1[L]$ and a matrix $\mathbf{G}$ were then calculated to perform (11). The former was obtained by $\mathbf{A}_1[L] = \mathbf{A}[L] - \mathbf{A}_0[L]$, where $\mathbf{A}_0[L]$ is obtained by $\mathbf{D}_{0L}\mathbf{A}[0]$. This CD matrix can be implemented as $\mathbf{D}_{z_1 z_2} = \mathbf{F}^{-1}\widetilde{\mathbf{D}}_{z_1 z_2}\mathbf{F}$, where $\mathbf{F}$ is the discrete Fourier matrix, $\widetilde{\mathbf{D}}_{z_1 z_2} = \mathrm{diag}\left(\exp\left(-\frac{j\beta_2}{2}\omega_0^2(z_2 - z_1)\right), \ldots, \exp\left(-\frac{j\beta_2}{2}\omega_{N-1}^2(z_2 - z_1)\right)\right)$, and $\omega_n$ is the angular frequency. $\mathbf{G}$ was calculated using (8) and (9) from the Tx signals $\mathbf{A}[0]$. For the calculation of $\mathbf{G}$, the spatial granularity $\Delta z$ was uniformly set to 0.5 km. 4.2e6 samples were used for PPE, and power profiles were averaged 50 times.

Fig. 2(a) shows simulation results for estimated longitudinal power profiles for a 50-km × 3-span link. A 1.0-dB attenuation was inserted at 75 km, and the launch power for each fiber span was set at 2, 4, and 0 dBm to test the capability of estimating non-uniform optical power levels. Note that, for clarity, the absolute optical power $\widehat{\mathbf{P}} = \widehat{\boldsymbol{\gamma}'}/\gamma$ is shown in the first vertical axis, assuming that $\gamma$ (= 1.30 W$^{-1}$km$^{-1}$) is known. A power profile of CM [18], [6] is also shown, for which the second vertical axis is used because CM does not estimate the true value of the signal power. The proposed linear LS (red) shows agreement with the theoretical line (black dashed), providing a reliable estimation of physical link parameters such as non-uniform fiber launch powers, fiber loss coefficients, the location of loss anomalies, and amplifier gains. In contrast, the power profile estimated by CM (blue) shows a smoothed characteristic and limited spatial resolution due to the convolution effect. Although CM provides a tendency of signal power variation, the estimated power deviates from the true power. Consequently, some calibration methods, as proposed in [21], are required for CM to correctly estimate the true physical parameters. Furthermore, CM exhibits a lower power level in the first span than in the third span despite that the true power in the first span is 2-dB higher than the third span. A similar observation can be found in [32]. This discrepancy is attributed to the fact that CM is largely dependent on the modulation format, while LS is not, as discussed in Remark 2.

Fig. 2(b) shows a magnified view around the inserted loss with various attenuation levels. The proposed linear LS accurately tracks these attenuation events, thus allowing the estimation of the attenuation levels. Notably, the method can detect even an attenuation as tiny as 0.2 dB, which is common in splice or connector losses. These findings illustrate that the linear least squares has a potential not only to test optical power levels but also to locate and estimate connector losses in a link, similar to OTDR.

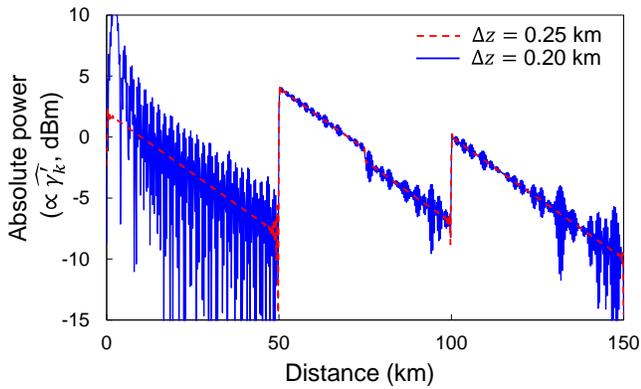

Fig. 3. Simulation results of PPE for $\Delta z$ = 0.25 and 0.20 km. Excessively fine spatial granularity inhibits stable PPE. 16QAM 128 GBd signals and $\beta_2$ = -21.6 ps$^2$/km were used. No noise and distortions were added.

*F. Ill-posedness and Limitation*

The simulation results above demonstrate the effectiveness of the linear least squares for PPE. However, there exists a fundamental limit on the performance of the LS. Fig. 3 shows estimated power profiles with spatial granularities of $\Delta z$ = 0.25 and 0.2 km, with no noise and distortion added. Although the estimation was stable at $\Delta z$ = 0.25 km or larger, the power profile collapsed at $\Delta z$ = 0.2 km, implying that there is an inherent limitation for stable PPE. This instability arises because the least squares problem (10) becomes ill-posed under finer $\Delta z$.

The ill-posedness of (10) is determined by a nonlinear perturbation matrix **G**, whose columns $\boldsymbol{g}_k$ form a basis for a nonlinear perturbed signal vector $\mathbf{A}_1[L]$. These columns are created by applying the CD, nonlinear operator, and residual CD to the Tx signals as shown in (9). When two of these columns are *close* to one another, the condition number of **G** grows, increasing the ill-posedness of the problem. For instance, when $\Delta z$ is small, linear independence of $\boldsymbol{g}_k$ and $\boldsymbol{g}_{k+1}$ is weakened because they are generated by similar operations: CD $\mathbf{D}_{0z_k}$ and $\mathbf{D}_{0z_{k+1}}$, the nonlinearity, and residual CD $\mathbf{D}_{z_k L}$ and $\mathbf{D}_{z_{k+1} L}$. In such a scenario, the matrix **G** possesses a large condition number, making the estimation prone to failure. The physical understanding is that signal waveforms at two closely positioned points $z_k$ and $z_{k+1}$ are similar, and the fiber nonlinearities they excite (and thus the optical power) are difficult to distinguish at the Rx. Similarly, when the CD effect (fiber CD coefficients or signal bandwidth) is small, $\boldsymbol{g}_k$ and $\boldsymbol{g}_{k+1}$ become more dependent, increasing the ill-posedness. In particular, in a dispersion managed (DM) link with dispersion compensating fibers, the estimation fails. In a DM link, fibers with opposite-sign CD coefficients coexist, and several columns in **G** completely match because multiple positions in a link share the same accumulated CD. This leads to identical signal waveforms at these positions, and excited nonlinearities cannot be distinguished. In such a case, the rank of **G** is reduced

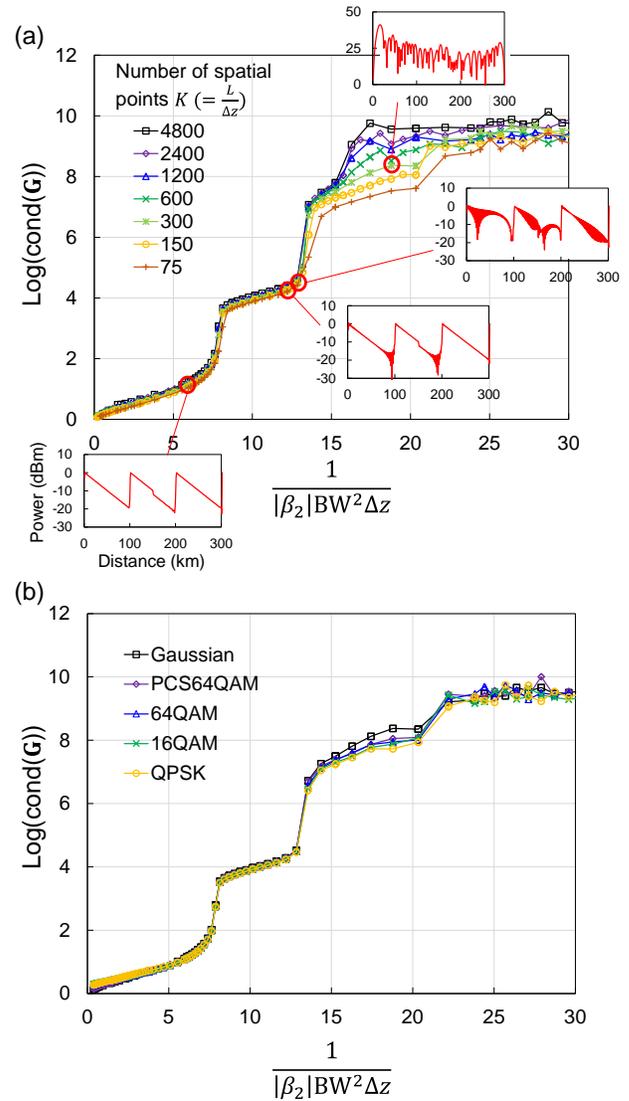

Fig. 4. Condition number of nonlinear perturbation matrix **G** as function of $\frac{1}{|\beta_2| \mathrm{BW}^2 \Delta z}$ (a) for various numbers of spatial points $K$ ( $= L/\Delta z$) with Gaussian signal format and (b) for various modulation formats with $K$ = 300 points. Insets in (a) are power profiles for $K$ = 300 ($L$ = 300 km and $\Delta z$ = 1 km.)

(the condition number is infinitely large), leading to the failure of the least squares estimation.

*G. Achievable Spatial Resolution*

From the discussion above, the achievable spatial resolution can be determined. Fig. 4(a)(b) shows the condition number of the matrix **G** obtained by simulations with various parameters. According to (8) and (9), **G** is dependent on CD coefficients, link distance $L$, Tx signal $\mathbf{A}[0]$, and spatial granularity $\Delta z$. Therefore, the simulations were conducted with all possible combinations of the following parameters:
- CD coefficients $\beta_2 \in \{-1, -2, \dots -41\}$ ps$^2$/km ($\beta_2(z) = const.$ and $\beta_3(z) = 0$ were assumed)
- Total distance $L \in \{75, 300, 1200\}$ km
- Modulation format $M \in$ {QPSK, 16QAM, 64QAM, PCS64QAM ($H$ = 4.347 bits, code rate = 0.826 [37]), Gaussian}
- Signal bandwidth BW $\in \{32, 64, 128, 256\}$ GHz

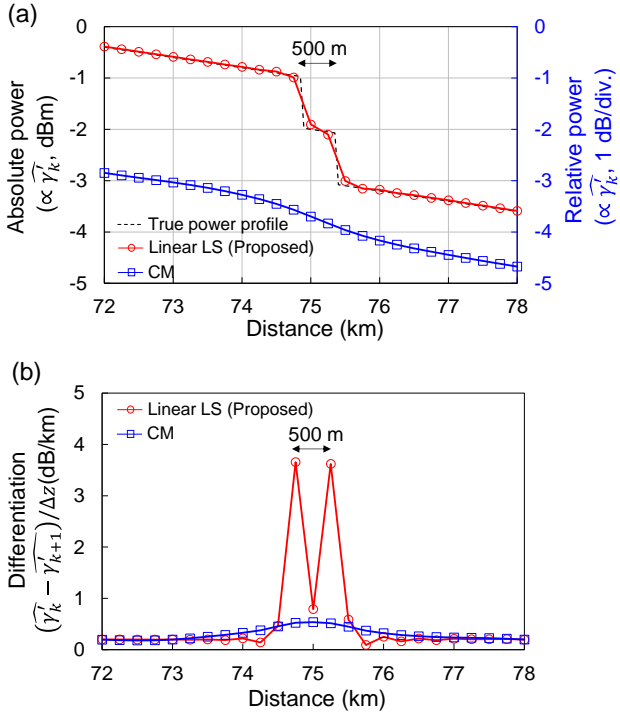

Fig. 5. (a) Simulation results of PPE with two 1.0-dB attenuations located 500-m apart. (b) Differentiation of (a) for detection and localization of anomaly loss events. $\Delta z$ = 0.25 km, 128-GBd signals, $\beta_2$= -21.6 ps$^2$/km.

- Spatial granularity $\Delta z \in \{0.25, 0.5, 1, 2\}$ km

Note that the Tx signal $\mathbf{A}[0]$ was shaped into a rectangular spectrum (the Nyquist limit,) and thus BW is equal to the signal symbol rate. Also, no noise and distortions were added to investigate a lower bound of the spatial resolution when PPE (insets in Fig. 4) was performed. For the horizontal axis, $\frac{1}{|\beta_2|\mathrm{BW}^2\Delta z}$ was chosen since the CD effect in a spatial step determines the condition number as discussed in the previous subsection. Fig. 4(a) shows the condition number for various numbers of spatial points ($K = \frac{L}{\Delta z}$) with the modulation format fixed to the Gaussian format, while Fig. 4(b) is for various modulation formats with $K$ fixed to 300. We found that all the curves almost form a unique line, which suggests that the horizontal axis $\frac{1}{|\beta_2|\mathrm{BW}^2\Delta z}$ is an effective metric to describe the evolution of the condition number across various modulation formats and spatial points. We also found that the number of spatial points $K$ and the modulation format slightly affect the condition number; however, they are not primary factors in determining the condition number within the range of stable estimation. As shown in the insets, the power profiles exhibit diverged characteristics as the condition number grows. The threshold of the condition number beyond which the PPE fails was observed to be approximately $10^{4.3}$. Correspondingly, $\frac{1}{|\beta_2|\mathrm{BW}^2\Delta z}$ should satisfy the following inequality to ensure stable estimation:

$$\frac{1}{|\beta_2|\mathrm{BW}^2\Delta z} < 12.84 \qquad (12)$$

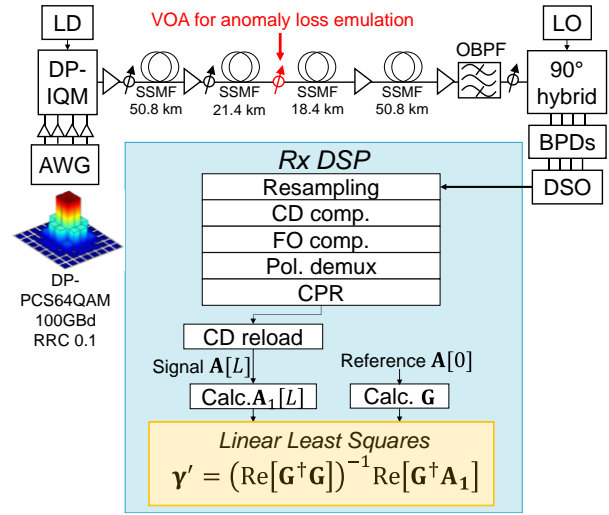

Fig. 6. Experimental setup and DSP-function blocks for linear-least-squares based longitudinal PPE.

In this work, the spatial resolution, denoted as SR, is defined as the minimal distance over which two consecutive power events can be distinguished. As shown in Fig. 5(a)(b), three measurement points are at least required to distinguish two loss events, which implies that SR > $2\Delta z_{limit}$ with $\Delta z_{limit}$ the achievable spatial granularity. Consequently, the lower bound of the spatial resolution of LS is expressed by transforming (12) as follows:

$$\mathrm{SR} > \frac{0.156}{|\beta_2|\mathrm{BW}^2} \qquad (13)$$
(for rectangular spectrum.)

Again, BW is equivalent to the signal symbol rate in a rectangular spectrum case. (13) implies that the achievable spatial resolution is improved by increasing fiber CD coefficients or the signal symbol rate. For instance, SRs of 1.76, 0.44, and 0.11 km are achieved for 64-, 128-, and 256-GBd signals, respectively, assuming $\beta_2$ = -21.6 ps$^2$/km. As a verification of (13), Fig. 5(a) shows the simulation results for power profiles with two consecutive 1-dB losses inserted 0.5 km apart. $\Delta z$, the symbol rate (=BW), and $\beta_2$ were set to 0.25 km, 128 GBd, and $\beta_2$= -21.6 ps$^2$/km, respectively. Fig. 5(b) is the differentiation of Fig. 5(a) $(\widehat{\gamma'_k} - \widehat{\gamma'_{k+1}})/\Delta z$, whose peaks indicate the location of loss events. For the proposed linear LS, two peaks were clearly observed, distinguishing two inserted losses with a spatial resolution of 0.5 km. However, with a finer $\Delta z$ such as 0.2 km (corresponding SR = 0.4 km), the power profile collapses as already observed in Fig. 3 (blue). These observations match the value calculated from (13), SR > 0.44 km. For CMs, only a single peak was observed, indicating the spatial resolution is more limited. This is due to the convolution effect inherent in CMs, as discussed in the Section II.C. Note that the simulation here assumed: (i) the signal had a rectangular spectrum, and (ii) $\Delta z_k$ was set uniformly. The extension to general spectra and non-uniform $\Delta z_k$ requires further analysis.

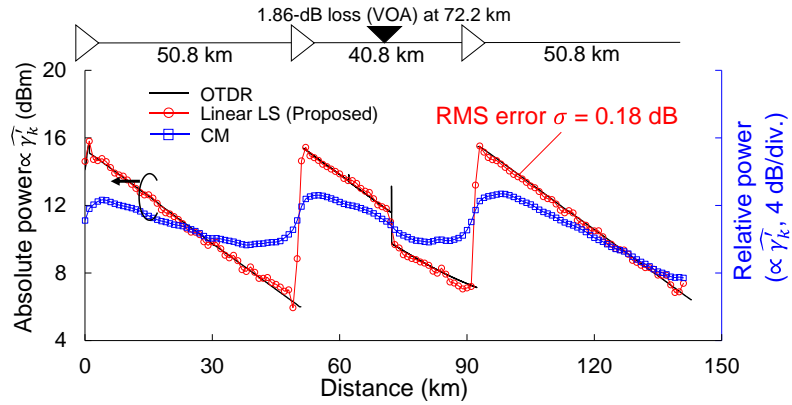

Fig. 7. Experimental results of PPE with proposed linear least squares (red) and correlation method (CM, blue) for 3-span link with 1.86-dB attenuation inserted at 72.2 km.

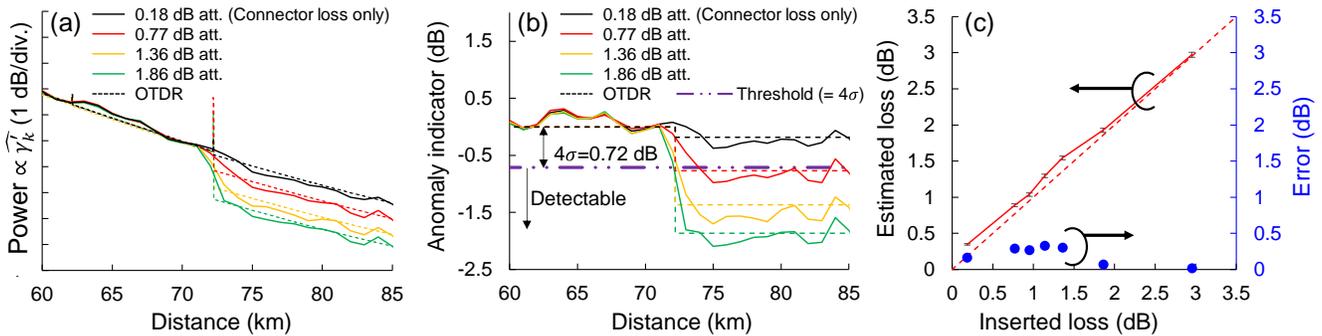

Fig. 8. (a) Estimated power profiles from 60 to 85 km with various VOA levels. (b) Anomaly indication by subtracting tilt (i.e., inherent fiber loss) from power profiles. Threshold for loss detection was set to $4\sigma = 4 \times 0.18 = 0.72$ dB. (c) Estimated loss as function of inserted loss (error bars with 100 power profiles).

## III. EXPERIMENT

### A. Experimental Setup

In this section, the experimental verification of the proposed method is presented under two conditions: (i) an ideal condition employing a single-channel transmission with high fiber launch power and (ii) practical conditions employing WDM with the optimal fiber launch power. The first scenario is ideal because PPE estimates the optical power by leveraging nonlinear SCI, and a high fiber launch power with less disturbance (i.e., cross channel interference, XCI) is preferred.

Fig. 6 shows the experimental setup for single channel transmission. The modulation format was PCS 64QAM with a roll-off factor of 0.1. The symbol rate was 100 GBd. The frequency response of the transmitter was estimated in advance and compensated for in the Tx DSP. The signal was emitted from a 4-ch 120-GSa/s arbitrary waveform generator (AWG), boosted by driver amplifiers, and converted to optical signals with a dual-polarization IQ-modulator (IQM). Tx and Rx lasers had a 1-Hz linewidth with a carrier frequency of 1547.31 nm. After amplified by an erbium-doped fiber amplifier (EDFA), the signal was launched into a 142.4-km 3-span standard single-mode fiber (SSMF) link with $\alpha = 0.180$ dB/km, $\beta_2 = -20.26$ ps$^2$/km, and $\gamma = 1.11$ W$^{-1}$km$^{-1}$. The fiber launch power was set to 15 dBm/ch. A variable optical attenuator (VOA) was inserted at 72.2 km to emulate the fiber anomaly loss. At the receiver side, out-of-band amplified spontaneous emission (ASE) noise was filtered out by an optical bandpass filter (OBPF). The optical signals were then detected by a 90° hybrid, balanced photodetectors (BPDs), and a 256-GSa/s digital sampling oscilloscope (DSO). In Rx DSP, resampling to 2 samples/symbols, CD compensation, frequency offset (FO) compensation, polarization demultiplexing, and carrier phase recovery (CPR) were applied. $\mathbf{A}[L]$ was then obtained by reloading the compensated CD to the signals after CPR. To perform the least squares estimation (11), a perturbation vector $\mathbf{A}_1[L]$ and a matrix $\mathbf{G}$ are required. These were computed from the transmitted signals $\mathbf{A}[0]$ as described in in Section II.E. In this experiment, we assumed that $\mathbf{A}[0]$ were known a priori. However, this does not mean that the PPE requires full pilot signals. This is because the transmitted signals can be recovered in the Rx thorough standard demodulation process. All the function blocks in the Rx DSP were conducted with 2 samples/symbol. The spatial step size $\Delta z$ was uniformly set to 1 km. Since this experiment used a dual-polarization transmission, the PPE algorithm was extended to the dual polarization case based on the Manakov equation as described in Appendix A.

The presence of noise such as ASE noise, phase noise, residual frequency offset, and XCI degrades the performance of PPE. However, such stochastically varying impairments can be mitigated by increasing the number of samples used for PPE or by averaging estimated power profiles. Details on these averaging numbers are described in each of the subsequent subsections.

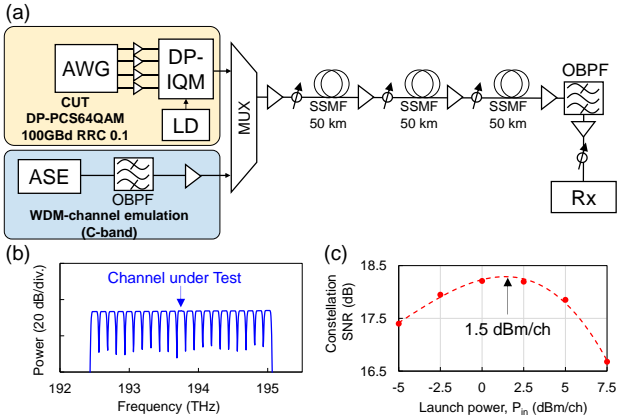

Fig. 9. Experimental setup for WDM transmission. (b) transmitted WDM spectra. (c) constellation SNR as function of fiber launch power. System optimal launch power was approximately 1.5 dBm/ch.

### B. Experimental Results Under Ideal Condition

First, the proposed method was tested in an ideal condition under a single-channel transmission with a fiber launch power of 15 dBm/ch. Fig. 7 shows the estimated longitudinal power profile over a 142.4-km 3-span link with a 1.86-dB attenuation inserted at 72.2 km. 2.5e6 samples were used for PPE and 100 power profiles were averaged. OTDR loss profiles are also shown for reference. The power profiles reproduced the simulation results in the previous section well for both the CM and proposed method. Although the CM captures the overall trend of the actual power, it deviates from the OTDR results and struggles to pinpoint the position of the loss anomaly. Thus, previous experimental demonstrations of CM [7], [8], [15], [21] relied on a normal state reference without loss anomaly and monitored the deviations from this reference to locate loss events. In contrast, the proposed linear LS aligns closely with the OTDR results, achieving an RMS error from OTDR of 0.18 dB and a maximum absolute error of 0.57 dB. Consequently, the inserted loss anomaly was clearly detected without using any normal state reference. Note that measurement dead zones of ± 1 km from the fiber ends were excluded from the error calculation.

To evaluate the detectable limit of loss anomalies, the VOA level was varied to 0.18, 0.77, and 1.36 dB. Fig. 8(a) presents a magnified view of the power profiles between 60 and 85 km, obtained by the proposed linear LS. These power profiles showed agreement with the OTDR results across all VOA levels. Notably, the estimated power profiles were highly reproducible even when VOA levels were varied as observed from the powers from 60 km to 70 km. To quantify the detectable limit, tilts of power profiles (i.e., inherent fiber losses) were subtracted from the power profiles, thereby revealing the amount of anomaly losses as depicted in Fig. 8(b). Considering that the RMS error of the power profiles was $\sigma = 0.18$ dB, the threshold for loss detection was set to $4\sigma = 4 \times 0.18$ dB $= 0.72$ dB. As the estimated power profile for a 0.77-dB loss anomaly (red) surpasses the 0.72-dB threshold, this 0.77-dB loss was effectively detected and localized. In addition, the amount of inserted losses can be estimated from Fig. 8(b) simply by averaging the estimated loss from the threshold-exceeding point (74 km) to the amplifier location (91 km). Fig. 8(c) shows the estimated loss as a function of the inserted loss. A total of 100 power profiles (without power profile averaging) were examined, and the estimated losses were highly stable with a standard deviation of < 0.03 dB and a maximum error of < 0.35 dB, demonstrating reliable estimation of values of loss anomaly.

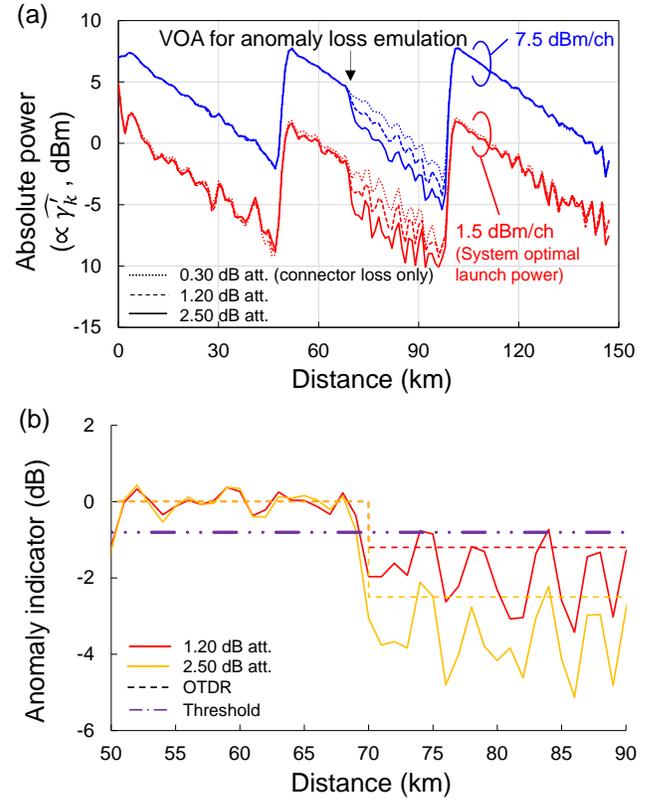

Fig. 10. Experimental results of (a) estimated power profiles under WDM conditions with various attenuation levels inserted. (b) Anomaly indication by subtracting tilt (i.e., inherent fiber loss) from power profiles at launch power of 1.5 dBm/ch.

### C. Experimental Results Under Practical Conditions

To investigate the performance of the proposed method under practical conditions, we conducted an additional experiment under WDM conditions with low fiber launch power. Fig. 9(a) shows the experimental setup. In this experiment, lasers with a 10-kHz linewidth were used for both Tx and Rx. The link comprises three spans of 50-km SSMFs, and WDM channels were emulated using an ASE source shaped by OBPFs. The channel under test (CUT) was set at 193.75 THz, and 20 adjacent WDM channels were arranged on a 125 GHz grid in the C-band (Fig. 9(b)). As shown in Fig. 9(c), the optimal fiber launch power for the system was approximately 1.5 dBm/ch. 8.1e5 samples were used to compute a power profile, and 50 profiles were averaged to enhance the SNR of power profiles. We determined these quantities so that power profiles sufficiently achieved convergence. All other conditions remained consistent with the previous experiment.

Fig. 10(a) presents the estimated power profiles under WDM condition with fiber launch power of 7.5 and 1.5 dBm/ch. A VOA for loss anomaly emulation was inserted at 70 km. At a launch power of 7.5 dBm/ch, a stable PPE was observed; however, the power profiles became noisier at the system optimal launch power (1.5 dBm/ch). In particular, estimated powers in the latter half of the spans were unstable. This instability occurs because an insufficient optical power stimulates weak nonlinearity, which is easily disrupted by link noise and distortions. The "information" of such a weak nonlinearity produced in fibers is challenging to detect at the Rx. In this experiment, the averaging effect (i.e., the number of samples used for PPE and the power-profile averaging) was sufficient to eliminate stochastic noise such as ASE noise and XCI. Static distortions such as transceiver imperfections were therefore the primary performance-limiting factor. At even lower launch powers, power profiles will experience more degradation.

Nevertheless, the power profiles were still clear enough to locate 1.2- and 2.5-dB attenuations, even at a launch power of 1.5-dBm/ch. Fig. 10(b) shows an anomaly indication by subtracting tilts from the estimated power profiles at 1.5-dBm/ch. Given that these power profiles contain more fluctuations compared to those under ideal conditions, we suggest an alternative method for detecting loss anomalies, where $\sigma$ is calculated as the RMS error from OTDR over locations prior to a loss event. For instance, $\sigma$ at 70 km is calculated as the RMS error from 51 to 69 km and was found to be 0.20 dB. Since the 1.2-dB attenuation level exceeds the $4\sigma$ threshold (0.80 dB), the loss event was successfully detected and located. The attenuation levels were estimated as an average from 71 km to 90 km, resulting in 1.9 dB and 3.6 dB for the actual attenuations of 1.2 dB and 2.5 dB, respectively.

We also found that to achieve $\sigma = 0.5$ dB (corresponding to a loss anomaly of 2.0 dB) at 70 km, approximately 5.0e4 and 4.0e5 samples in total were required for launch powers of 7.5 and 1.5 dBm, respectively. In scenarios with more noise such as long-haul transmissions, a greater number of samples will be necessary to maintain the same detection threshold. It can be inferred that a two-fold increase in stochastic noise doubles the samples required to detect the same loss anomaly. Therefore, in long haul systems where optical noise is dominant, the OSNR reduction of 3 dB also doubles the required sample size.

## IV. Conclusion

This paper presented a linear least squares method for Rx-DSP-based fiber-longitudinal PPE, which estimates the true value of nonlinear phase $\gamma'(z) = \gamma(z)P(z)$. We first showed that the estimation of $\gamma'(z)$, which is typically considered a nonlinear least-square problem, can be reduced to a simple linear least-square problem under the first-order regular perturbation approximation. As a result, this method finds the global optimum of least squares estimation, ensuring high measurement accuracy and spatial resolution. These characteristics of the method were validated in both simulations and experiments.

Simulations revealed that the method accurately aligns with theoretical powers, facilitating precise estimation of physical parameters throughout a link, such as fiber launch power levels, fiber loss coefficients, the amount and location of loss anomalies, and amplifier gains. Even a 0.2-dB loss anomaly was successfully located.

The fundamental performance limit of the least-squares-based method was also discussed in terms of the ill-posedness of PPE. This ill-posedness of PPE, making PPE prone to failure, intensifies when the CD effect in a tested link is weak. This tendency was evaluated quantitatively and comprehensively by assessing the condition number of the nonlinear perturbation matrix **G** for various link parameters. Consequently, the achievable spatial resolution was shown to be proportional to $1/\beta_2 \mathrm{BW}^2$.

In experiments, the method was first validated under ideal conditions with a high fiber launch power and single channel transmission. The estimated power profiles for 50 km × 3 spans achieved an RMS error from OTDR of 0.18 dB, successfully locating a loss anomaly as tiny as 0.77 dB, common in splice and connector losses. These results demonstrate that PPE operates similarly to OTDR at its maximum performance limit.

Furthermore, the performance of PPE was also investigated under practical conditions with system optimal fiber launch powers and WDM channels. Although the SNR of power profiles decreases with a reduction in optical power due to insufficient fiber nonlinearity, the estimated power profiles were still clear enough to locate loss anomalies even with system optimal launch powers.

Performance enhancement, the evaluation under long-haul transmissions, and the impact of noise and distortions require further studies and constitute the scope for future research.

## Appendix A

### Extension to Dual Polarization

In the case of dual polarization, one should consider inter-polarization nonlinearity to correctly estimate the true value of $\gamma'(z)$. To do so, the least squares formulation in (10) should be based on the Manakov equation [38]. By vertically stacking x- and y-polarization signal vectors, the cost function in (3) becomes:

$$I \simeq \left\| \begin{bmatrix} \mathbf{A}_x[L] \\ \mathbf{A}_y[L] \end{bmatrix} - \begin{bmatrix} \mathbf{A}_x^{ref}[L] \\ \mathbf{A}_y^{ref}[L] \end{bmatrix} \right\|^2 \quad (14)$$

Then, both $\mathbf{A}_{x/y}$ and $\mathbf{A}_{x/y}^{ref}$ are approximated by using RP1 such as $\mathbf{A}_{x/y}[L] = \mathbf{A}_{0,x/y}[L] + \mathbf{A}_{1,x/y}[L]$, where $\mathbf{A}_{0,x/y}$ is a linear term, and $\mathbf{A}_{1,x/y}$ is a first-order nonlinear term. If the sampling rate satisfies the Nyquist theorem for linear terms, $\mathbf{A}_{0,x/y} = \mathbf{A}_{0,x/y}^{ref}$ holds; then, the cost function is reduced to a comparison of the nonlinear terms only:

$$I \simeq \mathbb{E}\left[\left\| \begin{bmatrix} \mathbf{A}_{1,x}[L] \\ \mathbf{A}_{1,y}[L] \end{bmatrix} - \begin{bmatrix} \mathbf{A}_{1,x}^{ref}[L] \\ \mathbf{A}_{1,y}^{ref}[L] \end{bmatrix} \right\|^2\right] \quad (15)$$

By using RP1, $\mathbf{A}_{1,x/y}^{ref}$ can be expressed in a matrix form such as $\mathbf{A}_{1,x}^{ref} = \mathbf{G}_x \boldsymbol{\gamma}'$, where $\boldsymbol{\gamma}' = [\gamma'_0, \dots, \gamma'_{K-1}]^T$. The $k$-th column of $\mathbf{G}_{x/y}$ is

$$(\mathbf{G}_{x/y})_k = -j\Delta z \mathbf{D}_{z_k L}\left[\left(\mathbf{A}_{0,x}^*[z_k] \odot \mathbf{A}_{0,x}[z_k]\right.\right.$$
$$\left.\left.+ \mathbf{A}_{0,y}^*[z_k] \odot \mathbf{A}_{0,y}[z_k] - \frac{3}{2}\bar{P}\right) \quad (16)\right.$$
$$\left.\odot \mathbf{A}_{0,x/y}[z_k]\right]$$

where $\odot$ denotes element-wise multiplication, and $\mathbf{A}_{0,x/y}[z_k] = \mathbf{D}_{0z_k}\mathbf{A}_{x/y}[0]$. The cost function can then further be transformed as follows:

$$I \simeq \left\|\begin{bmatrix}\mathbf{A}_{1,x}\\\mathbf{A}_{1,y}\end{bmatrix} - \begin{bmatrix}\mathbf{G}_x\\\mathbf{G}_y\end{bmatrix}\boldsymbol{\gamma}'\right\|^2. \quad (17)$$

This can also be solved by linear least squares. By denoting $\mathbf{A_1} = \begin{bmatrix}\mathbf{A}_{1,x}\\\mathbf{A}_{1,y}\end{bmatrix}$ and $\mathbf{G} = \begin{bmatrix}\mathbf{G}_x\\\mathbf{G}_y\end{bmatrix}$, the real-valued solution is expressed similarly as (11):

$$\widehat{\boldsymbol{\gamma}'} = (\text{Re}[\mathbf{G}^\dagger \mathbf{G}])^{-1}\text{Re}[\mathbf{G}^\dagger \mathbf{A_1}]. \quad (18)$$

Note that, in this case, optical powers are estimated as $\widehat{\boldsymbol{P}} = \frac{9}{8}\frac{\widehat{\boldsymbol{\gamma}'}}{\gamma}$.

## APPENDIX B

### DERIVATION OF (11)

The cost function (10) is expanded as follows:

$$I = \|\mathbf{A}_1\|^2 + \boldsymbol{\gamma}'^T \mathbf{G}^\dagger \mathbf{G} \boldsymbol{\gamma}' - \boldsymbol{\gamma}'^T \mathbf{G}^\dagger \mathbf{A}_1 - \mathbf{A}_1^\dagger \mathbf{G} \boldsymbol{\gamma}'. \quad (19)$$

Differentiating (19) with respect to $\boldsymbol{\gamma}'$ yields

$$\frac{\partial I}{\partial \boldsymbol{\gamma}'} = (\mathbf{G}^\dagger \mathbf{G} + (\mathbf{G}^\dagger \mathbf{G})^T)\boldsymbol{\gamma}' - \mathbf{G}^\dagger \mathbf{A}_1 - \left(\mathbf{A}_1^\dagger \mathbf{G}\right)^T$$
$$= 2\text{Re}[\mathbf{G}^\dagger \mathbf{G}]\boldsymbol{\gamma}' - 2\text{Re}[\mathbf{G}^\dagger \mathbf{A}_1], \quad (20)$$

where formulas for a real vector $\mathbf{x}$ such as $\frac{\partial}{\partial \mathbf{x}}\mathbf{x}^T \mathbf{A}\mathbf{x} = (\mathbf{A} + \mathbf{A}^T)\mathbf{x}$ and $\frac{\partial}{\partial \mathbf{x}}\mathbf{x}^T\mathbf{a} = \frac{\partial}{\partial \mathbf{x}}\mathbf{a}^T\mathbf{x} = \mathbf{a}$ are used. Solving $\frac{\partial I}{\partial \boldsymbol{\gamma}'} = 0$ gives (11).